\documentclass[a4paper]{article}

\usepackage{INTERSPEECH2022}
\usepackage{booktabs,subfigure,multirow}

\usepackage{hyperref}

\title{VQTTS: High-Fidelity Text-to-Speech Synthesis with Self-Supervised VQ Acoustic Feature}
\name{Chenpeng Du, Yiwei Guo, Xie Chen, Kai Yu$^*$\thanks{* Corresponding author.}}

\address{
MoE Key Lab of Artificial Intelligence, AI Institute\\
X-LANCE Lab, Department of Computer Science and Engineering\\
Shanghai Jiao Tong University, Shanghai, China}
\email{\{duchenpeng, cantabile\_kwok, chenxie95, kai.yu\}@sjtu.edu.cn}

\begin{document}

\maketitle
\begin{abstract}
The mainstream neural text-to-speech(TTS) pipeline is a cascade system, including an acoustic model(AM) that predicts acoustic feature from the input transcript and a vocoder that generates waveform according to the given acoustic feature. However, the acoustic feature in current TTS systems is typically mel-spectrogram, which is highly correlated along both time and frequency axes in a complicated way, leading to a great difficulty for the AM to predict. Although high-fidelity audio can be generated by recent neural vocoders from ground-truth(GT) mel-spectrogram, the gap between the GT and the predicted mel-spectrogram from AM degrades the performance of the entire TTS system. In this work, we propose VQTTS, consisting of an AM txt2vec and a vocoder vec2wav, which uses self-supervised vector-quantized(VQ) acoustic feature rather than mel-spectrogram. We redesign both the AM and the vocoder accordingly. In particular, txt2vec basically becomes a classification model instead of a traditional regression model while vec2wav uses an additional feature encoder before HifiGAN generator for smoothing the discontinuous quantized feature. Our experiments show that vec2wav achieves better reconstruction performance than HifiGAN when using self-supervised VQ acoustic feature. Moreover, our entire TTS system VQTTS achieves state-of-the-art performance in terms of naturalness among all current publicly available TTS systems.
  
\end{abstract}

\vspace{-0.10cm}
\section{Introduction}

Text-to-speech (TTS) synthesis is a process that transforms a transcript into its corresponding speech. Compared with traditional statistical parametric speech synthesis \cite{spss}, neural TTS model \cite{wavenet, tacotron} based on deep neural network shows a better performance. The mainstream neural text-to-speech(TTS) pipeline is a cascade system, including an acoustic model(AM) that predicts acoustic feature from the input transcript and a vocoder that generates waveform according to the given acoustic feature. Two well-known AMs are Tacotron 2 \cite{tacotron2} based on encoder-attention-decoder architecture and FastSpeech 2 \cite{fastspeech2} based on Transformer blocks. As for the vocoders, generative adversarial network (GAN) \cite{gan} based vocoders, such as multi-band MelGAN \cite{multiband_melgan} and HifiGAN \cite{hifigan}, are widely used for their high quality of speech and fast generation speed. Another important type of vocoders is neural source-filter model \cite{nsf, nhv} based on the mechanism of human voice production.

However, the acoustic feature in current models is typically mel-spectrogram, which is highly correlated along both time and frequency axes in a complicated way, leading to a great difficulty for the AM to predict. Although high-fidelity audio can be generated by the neural vocoders from ground-truth(GT) mel-spectrogram, the gap between the GT mel-spectrogram and the predicted one from AM degrades the performance of the entire TTS system. 

Generally, there are two approaches to alleviate this problem. The first one is to provide stronger conditions for the AM in addition to the input transcript, such as prosody and linguistic feature. In this way, the AM could be more certain in acoustic feature prediction and often provides a better speech quality. The prosodies in different granularities have been provided to TTS in the literature. For example, \cite{pl1, phone_level_3dim, pl2} uses phoneme-level prosodies, \cite{wl1} uses word-level prosodies and \cite{hier1, hier2} uses hierarchical prosodies. The effectiveness of VQ-VAE is also explored in the related literature \cite{vqvae_f0,vqvae_pros}. In addition, \cite{graphtts, graphspeech} introduces syntactic graph and \cite{ac_wrd,pngbert} introduces word embeddings to TTS models.
Another approach to address the problem is to exploit better training criterion. The most common training criterion for the AM is the L1 or L2 loss, which assumes the distribution of the acoustic feature is unimodal. However, the real distribution is much more complicated. Hence, some research uses normalizing flow \cite{flow} in AM, such as FlowTTS \cite{flowtts} and GlowTTS \cite{glowtts}. The normalizing flow transforms the data distribution into a known simple distribution and is optimized via maximum log likelihood. However, the flow models should be carefully designed to ensure invertibility, which greatly restricts the capabilities of such models.

In this work, we propose VQTTS, consisting of an AM txt2vec and a vocoder vec2wav, which uses self-supervised vector-quantized(VQ) acoustic feature rather than mel-spectrogram. In particular, txt2vec basically becomes a classification model rather than a traditional regression model while vec2wav uses an additional feature encoder before HifiGAN generator for smoothing the discontinuous quantized feature. 
  Instead of predicting the complicated mel-spectrogram which is highly correlated along both time and frequency axes, txt2vec only needs to consider the correlation along time axis in feature prediction, which narrows the gap between GT and predicted acoustic feature dramatically. 
  Our experiments show that vec2wav achieves better reconstruction performance than HifiGAN when using self-supervised VQ acoustic feature. Moreover, our entire TTS system VQTTS achieves state-of-the-art performance in terms of naturalness among all current publicly available TTS systems.

\begin{figure*}[t]
  \centering
  \subfigure[txt2vec]{
    \includegraphics[width=0.51\linewidth]{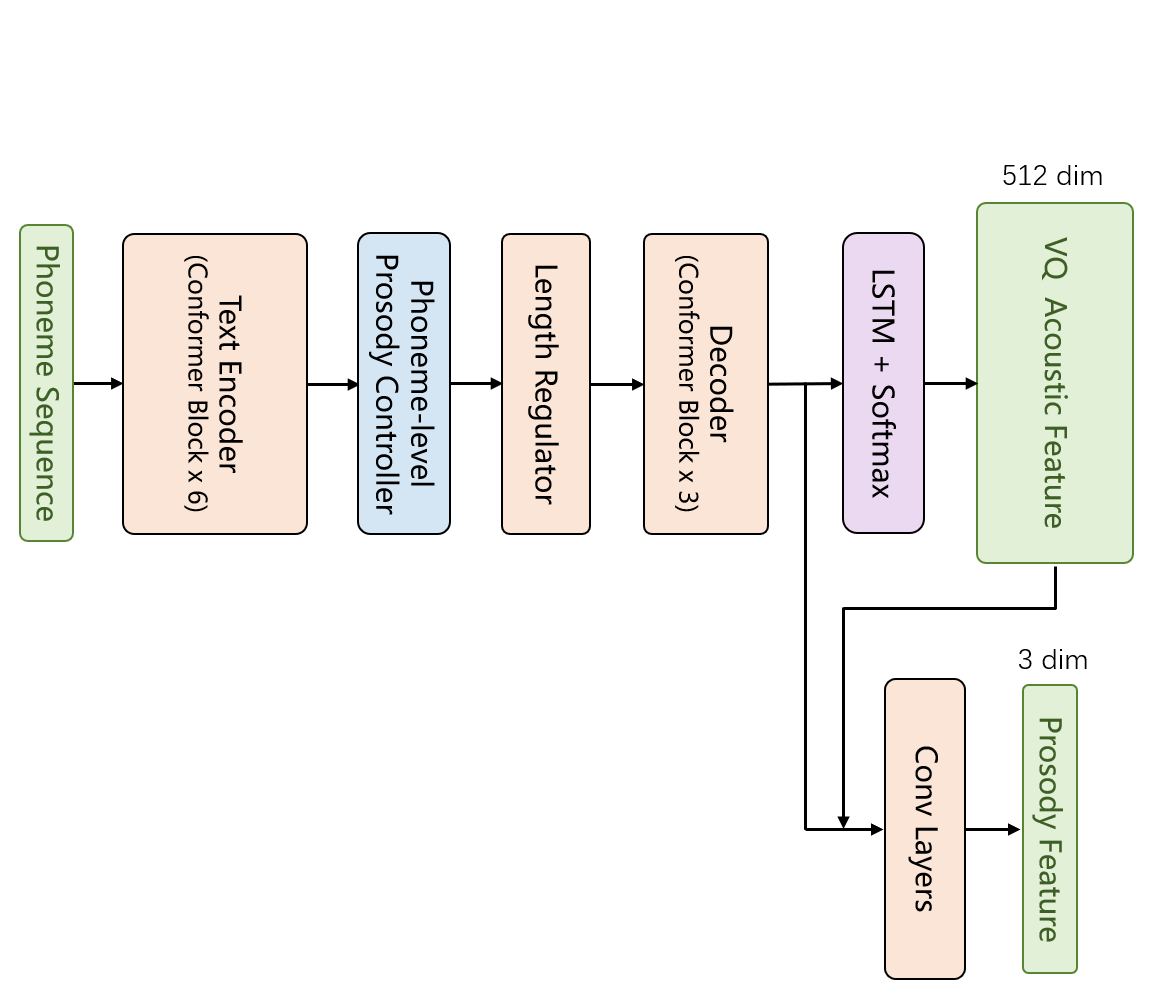}
    \label{txt2vec}
  }
  \subfigure[vec2wav]{
    \includegraphics[width=0.44\linewidth]{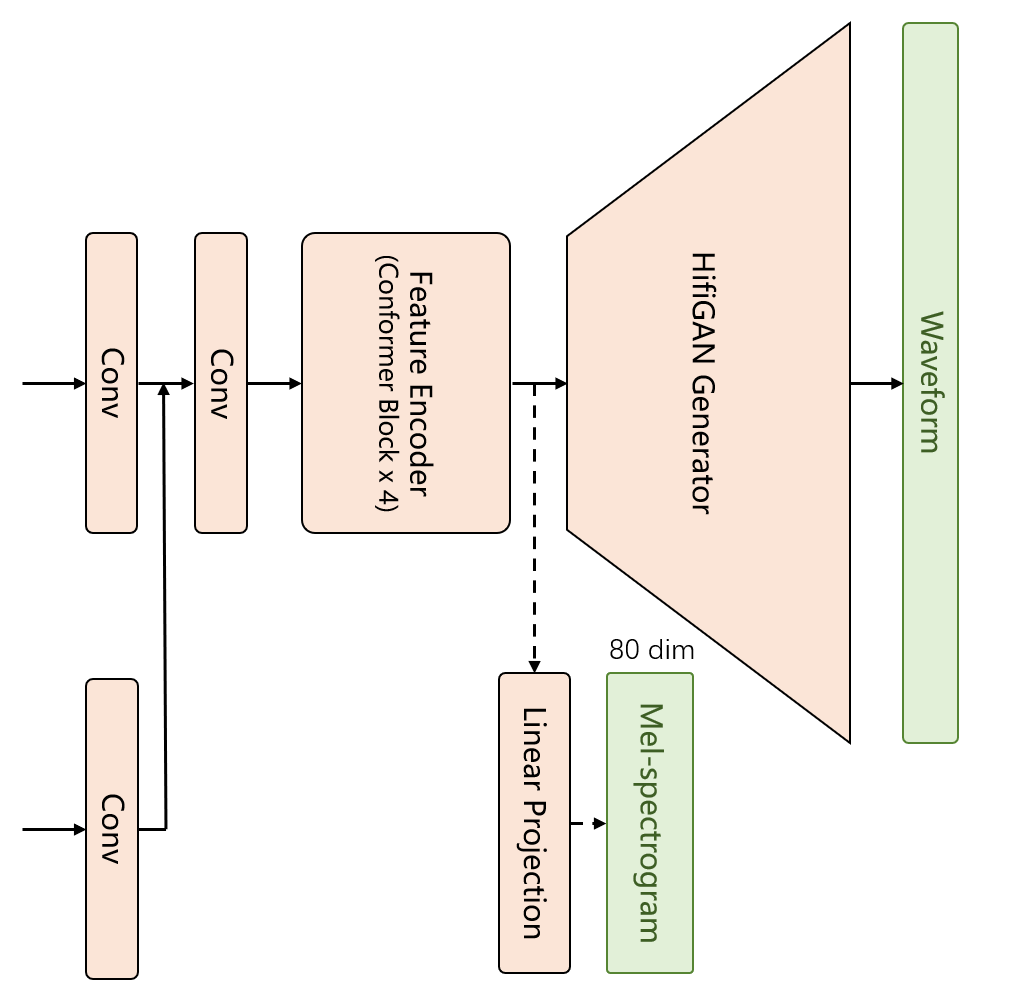}
    \label{vec2wav}
  }
  \caption{Model architecture of VQTTS, consisting of txt2vec and vec2wav. The two parts are connected with VQ acoustic feature together with prosody feature.}
  \label{fig:VQTTS}
  \vspace{-0.3cm}
\end{figure*}

\section{Self-Supervised VQ Acoustic Feature}

Recently, the acoustic features extracted by deep neural networks have been found to be superior to traditional acoustic features in automatic speech recognition (ASR). These types of neural networks are typically trained with only speech data in a self-supervised manner. They take the raw speech waveform $\mathcal{X}$ as input and generate the features $\mathcal{Z}$ that represents the characteristics of the speech segments. For example, wav2vec \cite{w2v} trains a multi-layer convolutional network optimized via a contrastive loss. Specifically, it tries to extract the features where we can predict several successive frames from the current and several previous frames. 

Later, vector quantization is applied to self-supervised feature extraction. Vq-wav2vec \cite{vqw2v} quantizes the acoustic feature $\mathcal{Z}$ to VQ acoustic feature $\mathcal{\hat{Z}}$ with gumbel-softmax or k-means. Then $\mathcal{\hat{Z}}$ is used to train a BERT \cite{bert} model for ASR initialization. In avoid of mode collapse where only a little amount of vectors in the codebook are actually used, vq-wav2vec divides the dimension of $\mathcal{\hat{Z}}$ into 2 groups and quantizes them separately. Based on that, wav2vec 2.0 \cite{w2v2} jointly trains the feature extractor and the BERT via a contrastive loss and a diversity loss that encourages the model to utilize more vectors in the codebook. 
HuBERT \cite{hubert} introduces another quantization strategy. Instead of training the codebook jointly with the feature extractor, HuBERT clusters the features with k-means in advance.

 In addition to the ASR task, self-supervised VQ acoustic feature has been also proven to be effective in a range of other tasks, such as voice conversion \cite{vqvc_nips, vqvc_is}, speech translation \cite{speech_translate} and speech separation \cite{speech_separate}.
 In this paper, we further investigate the use of VQ acoustic feature in the TTS task.

\vspace{-0.15cm}
\section{VQTTS}

VQTTS uses self-supervised VQ acoustic feature for speech synthesis. It has been found in the literature \cite{vqvc_is,vqvc_nips} that waveform reconstruction from VQ acoustic feature needs additional prosody feature. Therefore, in this work, we use 3 dimensional prosody feature, including log pitch, energy and probability of voice(POV)\cite{kaldi_pitch}. The prosody features are then normalized to zero means and unit variances. For simplicity, we abbreviate the combination of the VQ acoustic feature and the 3 dimensional prosody feature to VQ\&pros in the following sections. VQTTS contains two parts, the acoustic model txt2vec which predicts VQ\&pros from input phoneme sequence and the vocoder vec2wav which generates waveform from VQ\&pros. Here we elaborate the two parts in the following two subsections respectively.

\subsection{txt2vec}

\vspace{-0.10cm}
\subsubsection{Model architecture}

Before training txt2vec, we label the phoneme-level(PL) prosodies for all phonemes in advance, which will be detailed in the next subsection. Here we first demonstrate the overall model architecture of txt2vec in Figure \ref{txt2vec}. The text encoder consists of 6 Conformer blocks \cite{conformer}, which encodes the input phonemes into hidden states $\textbf{h}$. The hidden states are then sent to a PL prosody controller which predicts the PL prosody labels and a duration predictor that predicts the duration for each phoneme. After that, we repeat the hidden states according to the corresponding phoneme durations as in FastSpeech 2. The decoder is composed of 3 Conformer blocks and its output is passed through an LSTM layer followed by a Softmax activation function for VQ acoustic feature classification. Then the decoder output and the VQ acoustic feature are further concatenated and passed into 4 convolution layers, each followed by layer normalization and a dropout layer, for prosody feature prediction. The phoneme duration and prosody feature are trained with L2 and L1 loss respectively while the PL prosody label and the VQ acoustic feature are trained with cross entropy loss. The overall training criterion is
\begin{equation}
\label{eq:t2v_loss}
\mathcal{L}_{\texttt{txt2vec}} = \mathcal{L}_{\texttt{PL\_lab}} + \mathcal{L}_{\texttt{dur}} + \mathcal{L}_{\texttt{VQ}} + \mathcal{L}_{\texttt{pros}}.
\end{equation} 

\subsubsection{Phoneme-level prosody labelling}

As mentioned earlier, we have 3 dimensional normalized prosody features $\textbf{p}$ and then we calculate their dynamic features $\Delta\textbf{p}$ and $\Delta^2\textbf{p}$. The total 9 dimensional prosody features $[\textbf{p}, \Delta\textbf{p}, \Delta^2\textbf{p}]$ are averaged over the frames within each phoneme, so that we can represent the prosody of each phoneme with one vector. Then we cluster all the PL prosody representations into $n$ classes with k-means and take the cluster index as the PL prosody label.

\begin{figure}[h]
\centering
\includegraphics[width=0.8\linewidth]{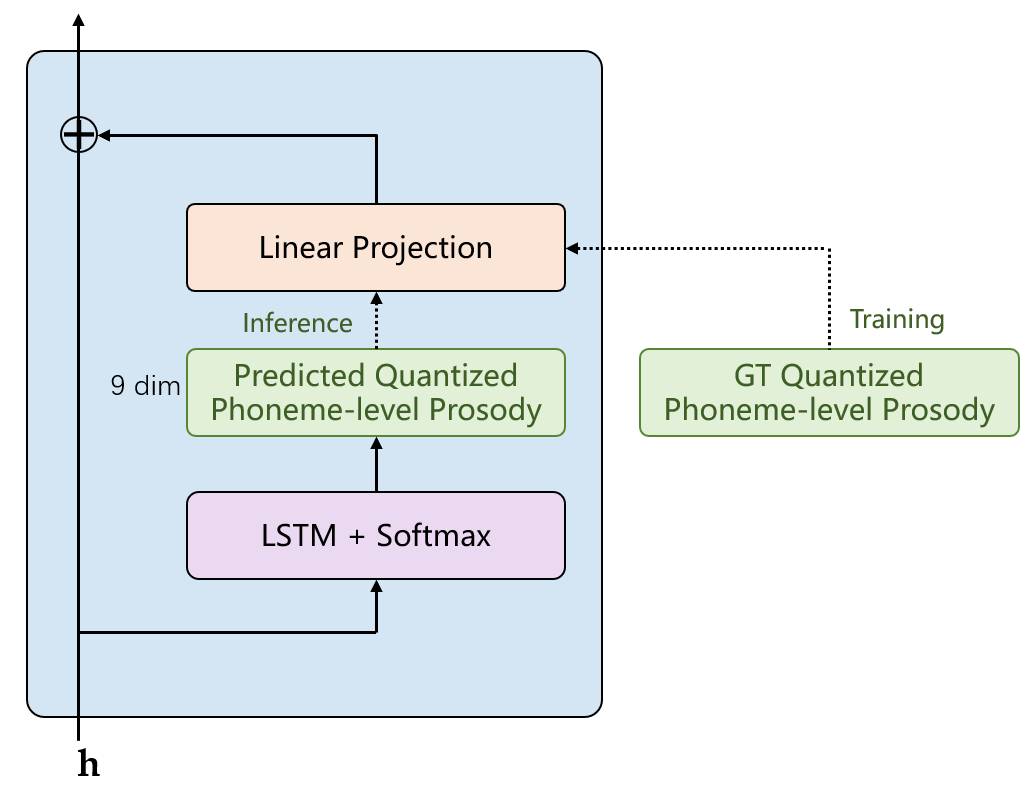}
\caption{The detail of phoneme-level prosody controller. The PL prosodies are quantized with the k-means.}
\label{fig:pl_prosody}
\end{figure}

\vspace{-0.25cm}

The architecture of the PL prosody controller is illustrated in Figure \ref{fig:pl_prosody}, which is trained to predict the PL prosody labels from the text encoder output $\textbf{h}$ with an LSTM. Then the quantized PL prosodies, i.e. the centers of the corresponding k-means clusters, are then projected and added to $\textbf{h}$ for controlling the following acoustic feature generation. Note that we use GT quantized PL prosodies in training and predicted ones in inference.

\vspace{-0.25cm}
\subsubsection{Beam search decoding}
There are two LSTMs in txt2vec, which are used for the autoregressive predictions of PL prosody label and VQ acoustic feature respectively. 
During training, both the LSTMs are conditioned on their inputs and the ground-truth previous outputs. During inference, we apply beam search decoding. In particular, the decoding starts with an all-zero vector \texttt{<sos>}. We denote the beam size as $k$ here. At each decoding step, we consider the top $k$ classes for all current hypotheses and take the results with the top $k$ probabilities as the new $k$ hypotheses. Compared with greedy search decoding that always selects the best result at each step based on the history, beam search decoding considers both the history and the future.

\vspace{-0.1cm}
\subsection{vec2wav}

\vspace{-0.05cm}
\subsubsection{Model architecture}
The model architecture of vec2wav is demonstrated in Figure \ref{vec2wav}. Both the VQ acoustic feature and the prosody feature are transformed with a convolution layer whose channels are 92 and 32 respectively and the kernel sizes are 5. The two outputs are then concatenated and passed to a convolution layer, a feature encoder, and a HifiGAN generator successively. The feature encoder here is designed for smoothing the discontinuous quantized acoustic feature. It contains 4 Conformer blocks, each of which uses 2 attention heads and 384 dimensional self-attention. The output of the HifiGAN generator is the corresponding waveform. The training criterion for HifiGAN is used for optimizing the vec2wav model.

\subsubsection{Multi-task warmup}

We find that vec2wav is hard to converge when we train the model from scratch with only HifiGAN loss. Therefore, we propose a multi-task warmup trick, which additionally uses a linear projection layer to predict the mel-spectrogram from the feature encoder output. Formally, we can write the training criterion during the warmup as
\begin{equation}
\label{eq:v2w_loss}
\mathcal{L}_{\texttt{vec2wav}} = \mathcal{L}_{\texttt{HifiGAN}} + \alpha \mathcal{L}_{\texttt{mel}}.
\end{equation} 
After the warmup, we remove the mel-spectrogram prediction task, which means the $\alpha$ is set to 0.

\section{Experiments and Results}

\subsection{Experimental setup}

We use LJSpeech \cite{ljspeech} dataset in our experiments, which is an English dataset containing about 24 hours speech recorded by a female speaker. We leave out 100 utterances for validation and 150 utterances for testing. 
All the speech data in this work is resampled to 16kHz for simplicity. We use a publicly available pretrained k-means-based vq-wav2vec model\footnote{\url{https://github.com/pytorch/fairseq/tree/main/examples/wav2vec}.} for VQ acoustic feature extraction. The frame shift of vq-wav2vec is 10ms and the number of possible VQ acoustic vectors is 21.5k.
The 3-dimensional prosody feature is extracted by Kaldi \cite{kaldi_pitch}. Audio samples are available online\footnote{\url{https://cpdu.github.io/vqtts}.}.


\subsection{Speech reconstruction with vocoders}
We train both the vec2wav and HifiGAN on the training set with VQ\&pros using Adam optimizer for 1 million steps. In vec2wav training, we set $\alpha$ to 60 at the first 200k training steps for warmup. A HifiGAN with mel-spectrogram is also trained for comparison.
Then we evaluate the performance of speech reconstruction on the test set given GT acoustic feature in both subjective and objective ways. In particular, we perform a mean opinion score(MOS) listening test where 15 listeners are asked to rate each utterance from 1 to 5 in terms of speech quality. Each listener is presented with 20 utterances randomly selected from the test set. For objective evaluations, we compute PESQ \cite{pesq} which measures the similarity between synthetic speech and the corresponding recordings. We also analyze the Gross Pitch Error (GPE) \cite{GPE} which calculates the proportion of frames whose pitch discrepancy in the recording and synthetic speech is less than 20\% among voiced frames. The results are listed in Table \ref{tab:reconstruct}.

\begin{table}[h]
\caption{Speech reconstruction performance of vocoders on the test set.}
\label{tab:reconstruct}
\centering
\begin{tabular}{cc|ccc}
\hline
\textbf{Feature} & \textbf{Method} & \textbf{MOS} & \textbf{PESQ} & \textbf{GPE}(\%)   \\ \hline  \hline
 - & Recording & 4.86$\pm$0.04 & - & - \\ 
Mel    & HifiGAN   & 4.68$\pm$0.04 & 3.60 & 0.79 \\ \hline \hline
VQ\&pros & HifiGAN   & 4.53$\pm$0.06 & 2.38 & 0.98 \\
VQ\&pros & vec2wav & \textbf{4.79$\pm$0.06} & \textbf{2.54} & \textbf{0.76}  \\ \hline 
\end{tabular}
\end{table}

 In the objective evaluations, it can be found that vec2wav
 can better reconstruct the recordings than HifiGAN with VQ\&pros. Also, we can see that the PESQ value of vec2wav is worse than that of the HifiGAN with mel-spectrogram. This is largely due to the information loss brought by quantization. However, a closer reconstruction does not indicate a better speech quality. Actually, the differences between the generated speech from vec2wav and the HifiGAN with mel-spectrogram are almost imperceptible. In the subjective listening test, vec2wav performs better than the HifiGAN with VQ\&pros and achieves comparable quality to the HifiGAN with mel-spectrogram. As for the HifiGAN with VQ\&pros, we can sometimes hear some undesired artifacts, which could be caused by the discontinuous quantized input feature.

\subsection{Naturalness of text-to-speech synthesis}

We train the entire text-to-speech system VQTTS in which txt2vec is optimized with Adam optimizer for 1000 epochs. The number of PL prosody clusters $n$ is set to 128 and the beam sizes in beam search decoding are set to 5 and 10 in PL prosody and VQ acoustic feature prediction respectively. 
Then we compare VQTTS with other current popular methods, including Tacotron 2, GlowTTS, FastSpeech 2 and the fully end-to-end TTS model VITS \cite{vits}. In the first three baseline systems, we use 80 dimensional mel-spectrogram as the acoustic feature and HifiGAN as the vocoder. Again, MOS listening test is conducted in the same way as the previous section for evaluating the naturalness of the synthetic speech. The results are shown in Table \ref{tab:tts} with 95\% confidence interval. 

\begin{table}[ht]
\caption{Evaluation for the text-to-speech synthesis systems.}

\vspace{-0.1cm}
\label{tab:tts}
\centering
\begin{tabular}{l|c|c}
\hline
\textbf{Setup} & \textbf{Feature} & \textbf{MOS} \\ \hline\hline 
Recording                & -        & 4.86$\pm$0.04 \\
GT Mel + HifiGAN         & Mel      & 4.68$\pm$0.04 \\
GT VQ\&pros + vec2wav    & VQ\&pros & 4.79$\pm$0.06  \\  \hline \hline
Tacotron 2 + HifiGAN     & Mel      & 3.67$\pm$0.05  \\ 
GlowTTS + HifiGAN        & Mel      & 3.72$\pm$0.05 \\
FastSpeech 2 + HifiGAN   & Mel      & 3.79$\pm$0.05 \\
VITS                     & -        & 4.62$\pm$0.04 \\ \hline 
VQTTS (txt2vec + vec2wav)& VQ\&pros & \textbf{4.71$\pm$0.05} \\ \hline 

\end{tabular}
\end{table}

\vspace{-0.2cm}
As is expected, quality degradation can be observed in all the cascade baseline TTS systems compared with speech reconstruction from GT mel-spectrogram. Although the fully end-to-end model VITS has a similar quality to VQTTS, it sometimes suffers from unnatural prosodies. The proposed VQTTS, however, generates high-fidelity and natural speech and shows little quality degradation compared with speech reconstruction from GT VQ\&pros. Moreover, VQTTS is a cascade TTS system, which is more flexible than the fully end-to-end TTS system VITS.

\subsection{Prosody Diversity in PL prosody hypotheses}

\begin{figure}[h]
\centering
\includegraphics[height=5cm, width=\linewidth]{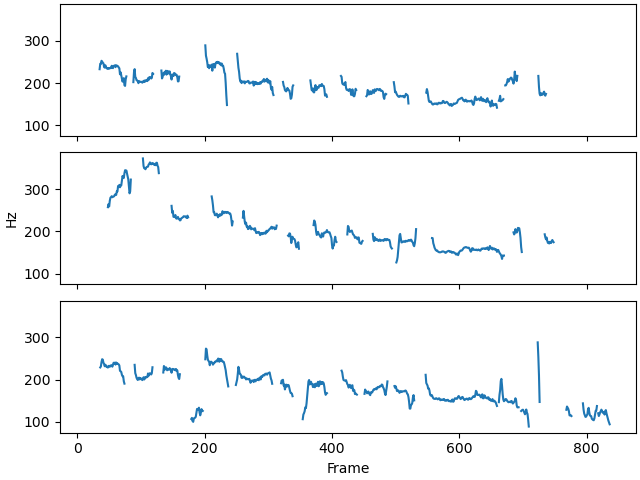}
\vspace{-0.5cm}
\caption{Pitch tracks of the synthetic speech with different prosodies.}
\label{fig:pitch}
\vspace{-0.5cm}
\end{figure}

Text-to-speech is a one-to-many mapping, since diverse prosodies are contained in speech in addition to the transcripts. VQTTS models the diversities with the PL prosody controller, which enables us to control the speech synthesis with different PL prosody hypotheses in beam search. Here we synthesize a sentence in the test set with 3 different prosody hypotheses and demonstrate their pitch tracks in Figure \ref{fig:pitch} where obviously we can see their differences.

\subsection{Decoding algorithm}

We explore the effectiveness of beam search decoding for both PL prosody label and VQ acoustic feature predictions. To this end, we exploit greedy search and beam search with a beam size of 5 and 10 in the two tasks separately. Here, VQ acoustic feature prediction is conditioned on the GT durations and PL prosody labels in order to make sure that the predicted feature is precisely aligned with the GT feature so that we can calculate the prediction accuracy. The results are presented in Table \ref{tab:ablation_prosody} and \ref{tab:ablation_vqfeat}. 

We can find that the accuracy in all setups is not so high because we have demonstrated the diversity of speech in the previous section. Despite that, the accuracy of beam search decoding is still slightly better than greedy search in both inference tasks. Also, the beam size of 5 is better in PL prosody label prediction while the size of 10 is better in VQ acoustic feature prediction.

\begin{table}[ht]
\caption{Prediction accuracy for phone-level prosody label.}
\vspace{-0.1cm}
\label{tab:ablation_prosody}
\centering
\begin{tabular}{l|c}
\hline
\textbf{Decoding Algorithm} & \textbf{Accuracy(\%)} \\ \hline
greedy search  & 12.12 \\ 
beam search (beam size = 5)   & \textbf{12.60} \\ 
beam search (beam size = 10)   & 12.46 \\ \hline 
\end{tabular}
\end{table}

\vspace{-0.2cm}
\begin{table}[ht]
\caption{Prediction accuracy for VQ acoustic feature.}
\vspace{-0.1cm}
\label{tab:ablation_vqfeat}
\centering
\begin{tabular}{l|c}
\hline
\textbf{Decoding Algorithm} & \textbf{Accuracy(\%)} \\ \hline
greedy search                 & 13.96 \\ 
beam search (beam size = 5)   & 14.09 \\ 
beam search (beam size = 10)   & \textbf{14.20} \\ \hline 
\end{tabular}
\end{table}

\vspace{-0.25cm}
\section{Conclusions}
In this paper, we propose VQTTS that utilizes self-supervised VQ acoustic feature rather than traditional mel-spectrogram, which dramatically narrows the quality gap between GT and predicted acoustic feature and consequently improves the performance of entire TTS system. The vocoder in VQTTS, vec2wav, uses an additional feature encoder for smoothing the discontinuous quantized input feature and achieves a better reconstruction performance than HifiGAN. We also find that diverse prosodies can be generated by the different PL prosody hypotheses in beam search decoding. Furthermore, beam search decoding performs better than greedy search in both PL prosody and VQ acoustic feature prediction.

\section{Acknowledgements}

State Key Laboratory of Media Convergence Production Technology and Systems Project (No.SKLMCPTS2020003), Shanghai Municipal Science and Technology Major Project (2021SHZDZX0102), National Natural Science Foundation of China (Grant No.92048205)

\bibliographystyle{IEEEtran}

\bibliography{mybib}

\end{document}